\documentclass[aps,prl,preprint,groupedaddress]{revtex4}
\usepackage{graphicx}
\usepackage{xcolor}
\usepackage{amssymb}
\usepackage{amsmath}
\usepackage{array}
\usepackage{lineno}
\usepackage[normalem]{ulem}
\definecolor{darkblue}{rgb}{0,0,0.5}
\definecolor{darkgreen}{rgb}{0,0.5,0}
\definecolor{darkred}{rgb}{.9,0,0}
\definecolor{purple}{rgb}{0.5,0,0.6}
\definecolor{orange}{rgb}{1,0.5,0}
\definecolor{grey}{rgb}{.6,.6,.6}
\definecolor{lightpink}{rgb}{1,0.7,0.75}
\definecolor{pink}{rgb}{1,0.4,0.58}
\definecolor{deeppink}{rgb}{1,0.08,0.58}

\begin{document}
\title{Design of a Single-Shot Electron detector with sub-electron sensitivity for electron flying qubit operation}
\author{D. C. Glattli, J. Nath, I. Taktak, P. Roulleau}
\affiliation{Universit\'e Paris-Saclay, CEA, CNRS, SPEC, 91191, Gif-sur-Yvette, France}

\author{C. B\"auerle}
\affiliation{Univ. Grenoble Alpes, CNRS, Grenoble INP, Institut N\'eel, 38000 Grenoble, France}

\author{X. Waintal}
\affiliation{Universit\'e Grenoble Alpes, CEA, IRIG-PHELIQS, Grenoble, France
}



\date{\today}
\begin{abstract}

The recent realization of coherent single-electron  sources in ballistic conductors let us envision performing time-resolved electronic interferometry experiments analogous to quantum optics experiments. One could eventually use propagating electronic excitations as flying qubits. However an important missing brick is the single-shot electron detection which would enable a complete quantum information operation with flying qubits. Here, we propose and discuss the design of a single charge detector able to achieve "in-flight" detection of electron flying qubits.  Its sub-electron sensitivity would allow the detection of the fractionally charged flying anyons of the Fractional Quantum Hall Effect and would enable the detection of anyonic statistics using coincidence measurements.

\end{abstract}

\pacs{73.23.-b,73.50.Td,42.50.-p,42.50.Ar} \maketitle

\section{\label{sec:level1} INTRODUCTION}

 Electron quantum optics \cite{BauerleReview,PhysStatSol,Glat2016} is a fast emerging field which aims at performing quantum  operations with electrons similar to those done with photons in quantum optics. 
 It opens a new way to embody electron-based quantum bits in condensed matter. 
 While the mainstream approach to quantum processing is based on localized two-level electronic systems ( spin states in semiconductor quantum dots, charge or flux states in superconducting circuits, etc.), here, delocalized electrons carry the quantum information. 
 They form \textit{ flying qubits } which propagate along the quantum modes of a conductor \cite{flying_quBits1,flying_quBits2,flying_quBits3}. 
 To be specific, a simple flying qubit quantum circuit is shown in Fig.1. 
 The conductor is a two-dimensional electron gas (2DEG) in high perpendicular magnetic field $B$ where the Quantum Hall Effect sets in. 
 In this regime, the bulk of the conductor is topologically insulating while chiral one-dimensional edge modes carry the current along the sample edge. 
 
 \begin{figure}
  \includegraphics[width=8.5cm,keepaspectratio=true,clip=true]{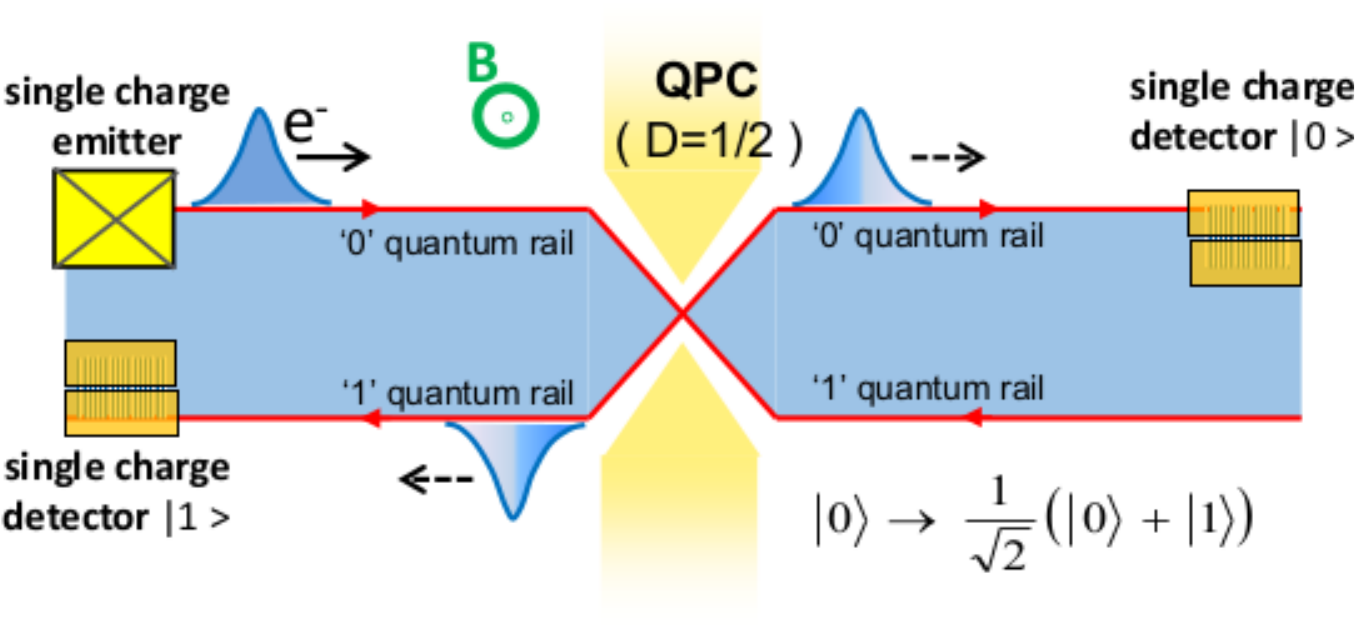}
  \caption{ Electronic flying qubit: 
  The conductor, a 2D electron gas  is shown in blue. 
  Under strong perpendicular magnetic field B, in the quantum Hall effect regime, the bulk of the conductor is insulating and current propagates along chiral edge channels (red lines with arrows). 
  Applying a short voltage pulse on a contact (yellow square with cross), a single charge is emitted in the upper left channel and its wave function is diffracted by a quantum point contact (QPC) which puts the electron in a quantum superposition of right transmitted and left reflected states. 
  Labeling the upper (lower) edges '0' and '1', this realizes the qubit state $\frac{1}{\sqrt{2}}(|0 \rangle + |1 \rangle ) $. 
  To perform single-shot quantum operation one needs "in-flight" detection of the flying qubit at each output as shown by the dashed orange box denoting the charge detectors at the upper right and lower left outputs.  These charge detectors are needed  to take advantage of a full operation of electronic flying qubits. }
  \label{detector}
\end{figure}
 A flying qubit is obtained by the on-demand injection of a single electron into an edge mode (here, called quantum rail "0" ), using a  single-electron source based on voltage pulses \cite{Dubo13,Jull14,coherentstates,Ivan97,lebePRB05,Keeling06}.
 In Fig.1 the flying electron is shown to propagate towards a Quantum Point Contact (QPC), a narrow constriction controlled by an electrostatic gate, which mixes the upper and lower counter-propagating edge modes. 
 The QPC can be viewed as the electronic analog of a photon beam-splitter as the quantum mixing puts the incoming charge in a quantum superposition of right transmitted and left reflected states. 
 Labelling the quantum state of an electron occupying rails "0" and "1" by $|0>$ and  $|1>$ respectively, one sees that the electron initially in the $|0>$ state ends up in the state $\frac{1}{\sqrt{2}}(|1>+|0>)$ for a QPC transmission $D=1/2$. 
 This performs the single qubit operation known as the Hadamard gate \cite{BauerleReview}. 
 Combining the QPC beam-splitter with phase delays \cite{Ji03,Roulleau08,Yama12} allows to realize all single qubit rotations on the Bloch sphere needed for quantum information processing. 
 A two-qubit gate for flying qubits  can be done by bringing two such qubits close together and use the Coulomb interaction to provide a conditional phase shift. Combining QPCs and 2-qubit gates will enable the realization of complex quantum algorithms \cite{BauerleReview}.

 Recent developments in Electron Quantum Optics, have provided various types of on-demand single electron sources \cite{Herm11,McNe11,Flet13,Ryu16,Ubbe14,Feve07,Dubo13}. 
 To perform a complete flying qubit quantum operation however, an important brick is missing: the single-shot detection of the flying qubits. In the example of Fig. 1 the measurement will tell us if the flying electron is occupying the output state  $|0>$ or $|1>$. 
 
 The aim of this paper is to discuss the design and efficiency of a charge detector appropriate for in-flight flying qubit detection. 
 
 The paper is organized as follows.  Section II discusses various approaches of Quantum Non Demolition (QND) and non-QND single charge detection and why detecting electron qubits on the fly is a technological challenge. In  section III we discuss the design of an original non-QND detector able to meet the requirement for good fidelity. In section IV we describe how coupling the non-QND detector to an electronic quantum interferometer can lead to QND detection.  We conclude by discussing the new opportunities enabled by this single-charge detector for quantum physics beyond the field of flying qubits.  

\section{\label{sec:level2} APPROACHES TO SINGLE CHARGE DETECTION}

We consider the "in-flight" detection of an electronic flying qubit by capacitive means using a charge detector coupled to the edge channel of the output lead. Here it is important to distinguish two important classes of detectors. 

The first type is a quantum non-demolition (QND) detector. 
This is a quantum detector able to record the charge information while not destroying the quantum state of the qubit. 
This detector must be fast and preferably must operate before quantum decoherence starts such that the flying qubit information can be re-used for further quantum information processing. 
A possible implementation has been discussed in \cite{BauerleReview} and a charge resolution of a few electrons has recently been achieved \cite{grenoble_2019}.

In the second type of detector, quantum demolition of the state occurs because the measuring time is so long that the flying qubit has lost coherence either due to detector back action or because of interaction with the environment. It is this type of non-QND flying qubit detector that we will first consider in  section III of this work. In section IV we will then show that QND detection can be reached if one puts this non-QND detector at the output of an electronic interferometer. 

\begin{figure}
  \includegraphics[width=8.5cm,keepaspectratio=true,clip=tue]{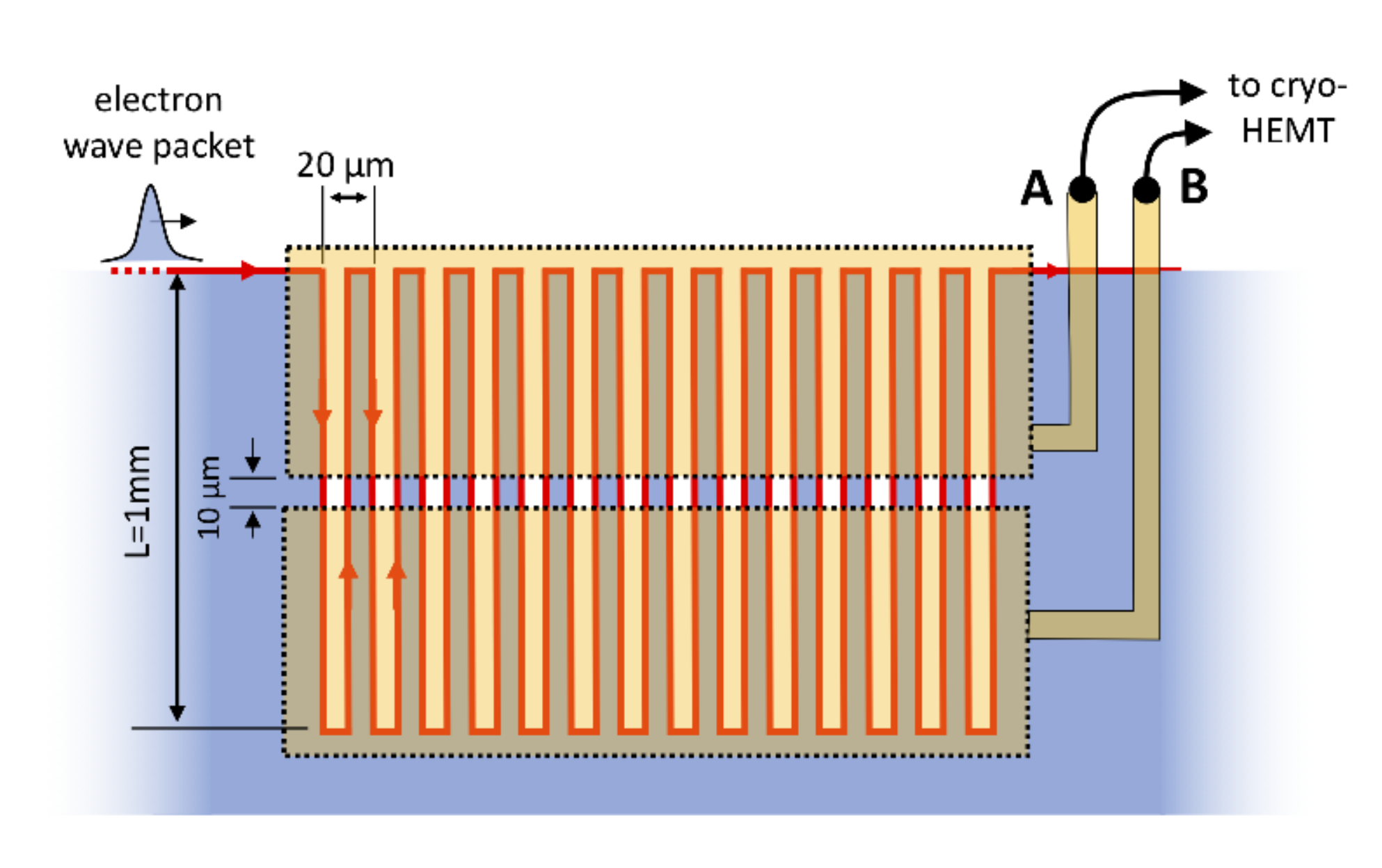}
  \caption{In-flight charge detector. The 2DEG electron gas, buried ~100nm under the  surface of the semiconductor heterostructure is shown in blue. The red line with arrows denotes the chiral edge channel which propagates at the 2DEG boundary. The boundary is patterned by etching in the form of mm long meanders with 20$\mu$m period. An electron flying qubit (shown as a blue "hat" in the upper left) with short wavepacket width ($\simeq 1\mu m$) enters the detection region where it is forced to meander under two metallic surface gates (shown in yellow) deposited on the chip surface. The image charge induced on the sensing electrodes A and B generates an oscillating voltage which is sent to a low noise cryogenic HEMT for amplification. About 75 periods of the meandering edge channel are necessary to achieve a sub electron detection.}
  \label{fig:detector}
\end{figure}

The detector has to detect single shot and with good fidelity the presence of a flying single charge in a given output lead, say rail "0" as sketched in Fig.1.  
Systems able to detect a localized charge with single electron accuracy have emerged in the $80's$ after the observation of Coulomb Blockade (CB) effects \cite{Fult87}. 
These electrometers require an ultra-small capacitor $C$, less than a femtoFarad,  such that the Coulomb charging energy $e^{2}/2C$ is larger than the subKelvin operating temperature. 
The small capacitance also provides a large detectable charging voltage $e/C$, more than a hundred of $\mu$V. Using submicron size metallic islands isolated from leads by tunnel barriers, Coulomb Blockade electrometers have shown a typical charge sensitivity of $10^{-5}\,e/$Hz$^{1/2}$ and allowed single electron detection with a few $\%$ accuracy in about $1 \mu s$ detection time \cite{ensslin_2007,marcus_2007,dzurak_2007}.

In recent experiments, such electrometers have been implemented in single-electron transport experiments where the propagating electron is captured after a propagation of several tens of micrometers with efficiencies well above 99 \% \cite{Freise2019,Takada2019}.
These electrometers, however, are not suitable for flying qubits as the charge moves very fast \cite{Kuma2011,Roussely2018}. 
For the specific case of the Quantum Hall effect at filling factor $\nu=1$ and a typical propagation velocity of $v\simeq3.10^4$\,m/s \cite{Kuma2011}, such electrometers with submicron size require to detect the charge in less than 30\,ps, a time far too short for detection with present technology.
 In contrast, a detection process lasting $1\,\mu$s would require a constant capacitive coupling along a 3\,cm long propagation length \cite{Kama14}. This results in a large sensing capacitance well above the femtoFarad needed to support Coulomb Blockade Effects.

Thus, contrasting with Coulomb blockade electrometers, the flying qubit charge detector must have a large sensing capacitance, several hundreds of fF, to maintain the coupling to the charge over a length of a cm. To detect and amplify the charge signal, cryogenic High Electron Mobility Transistors (HEMTs), as those designed for low noise microwave amplifiers, are good candidates. 
They offer an input capacitance of a fraction of picoFarad with very low input voltage noise at cryogenic temperature, typically 0.2\,nV/$\sqrt{Hz}$ white noise above few MHz detection frequency. 
For a typical detector input capacitance $C_{in}$ of 1 pF (see table 1) a single charge gives an input signal $V_{s}=e/C_{in}$ of 0.16\,$\mu$V. This is detectable by the HEMT in 1\,$\mu$s. 
However, this estimation only considers the white noise of the HEMT and not the extrinsic noise due to fluctuating charges in the HEMT electron channel. 
With $1/f$ type spectral noise density, the extrinsic noise overcomes the white shot noise for frequencies below a few MHz. 
Thus, after a detection time longer than a fraction of $\mu$s, integrating $V_{s}$ is ineffective and single-charge accuracy is not reachable. 

In the next section, we discuss the design of a new kind of capacitive detector
- the meander-capacitive detector - that can overcome the above limitations by converting the transient capacitive signal into an oscillatory signal in the $\sim 10-100\,$MHz frequency range.


\begin{widetext}
\begin{table}[h]
    \centering
    \begin{tabular}{||l|c|c||}
    \hline
    Characteristic   &  Definition & Estimate  \\
    \hline
    Pulse velocity   & $v$        &  $3.\ 10m.s^{-1}$ \\
    Number of meanders & $N$      &  $75$ \\
    Length of meanders & $L$      &  $1 mm$ \\
    Gap between gates & $G$  &  $10 \mu m$ \\
    Distance 2DEG surface & $d$ &  $100 nm$ \\
    Distance between meanders & $w$  & $20 \mu V$ \\
    HEMT noise level   & $S_{HEMT}$    &  $0.2nV/\sqrt{Hz}$ \\
    HEMT gain       & $g$   &  5-8  \\
    \hline
    Signal frequency  & $f = v/L$  &  $15 MHz$   \\
    Measurement time  & $\tau=NL/v$ &  $4.5 \mu s$ \\
    Voltage noise     & $V_{Noise} = S_{HEMT}/\sqrt{\tau}$ &  $0.093 \mu V$ \\
    Gate capacitance  & $C_{AB}(d,G,w)$  & $250 fF$   \\
    Total capacitance & $C_{in}=C_{AB} +C_{HEMT}+ C_{access}$  & $480 fF$  \\
    Voltage signal    & $V_{s}= e /C_{in}$ & $0.33 \mu V$ \\
    SNR               & $SNR = V_s/V_{Noise}$ &   3.6   \\
    \hline
    \end{tabular}
    \caption{Summary of the various estimates for the meander single charge detector}
    \label{tab:my_label}
    
\end{table}
\end{widetext}

\section{\label{sec:level3} The Meander-capacitive single electron single shot detector}
We now turn to the proposal of this article to build a single-shot single electron detector. We propose to overcome the $1/f$ noise issue by forcing the charge signal $e/C_{in}$ discussed above to oscillate at a frequency well above the frequency range where the HEMT is dominated by the $1/f$ noise such that integration of the signal occurs in the gaussian white noise regime. 

To do so, a known approach is to make the input capacitance oscillating so as to convert the D.C. signal into an A.C. signal. This is the vibrating-reed electrometer strategy used about 60 years ago for sensitive electrometers, dosimeters, and ph-meters like the  XL7900 tube by Philips. Here, we propose another original approach of this idea: taking advantage of the several cm-long path, we guide the flying charge along  edge channels meandering below two sensing capacitor electrodes.  


A schematic view of the detector is shown in Fig.\ref{fig:detector}.
The 2DEG edge is patterned into a meander line that considerably enlarge the length of the path of the single-electron excitation. Two capacitive gates are placed on top of the meander line. When the single electron excitation is underneath the first gate, it creates a mirror image there, and when it is underneath the second gate, the mirror image changes to the second gate. This creates an oscillating current of amplitude $i_{in}\simeq ef$ at a  frequency $f=v/(2L)$  through the impedance $Z$ linking the two capacitive gates, where $L$ corresponds to the length of one meander and $v$ to the velocity of the excitation.  
The resulting voltage $V_{s}=Z\,i_{in}$ is the signal to be amplified and detected.

First, we discuss the typical dimensions and the sensitivity of the proposed device. 
The detector is tailored in a 2DEG by etching a meander-line pattern with N=75 repetitions.

Each meander has a length of $ L=\,1\,$mm, a width of $W=$\,10\,$\mu$m and a spacing of $S=$ 10\,$\mu$m.
The two sensing metallic electrodes A and B  of size 1.5\,mm$\,\times\,$0.5\,mm are deposited on top of the meanders.
The distance between the electrodes and the two-dimensional electron gas is typically  $d\simeq 100$\,nm. For a gap  $G=10\mu$m separating the two sensing electrodes, their mutual capacitance is found to be $C_{AB}=250$ fF from 3D electrostatic numerical calculation.

Assuming the impedance $Z$ dominated by the gate and HEMT capacitance, the charge passing alternatively under electrode A and B creates an oscillating voltage of amplitude $V_{s}\simeq e/C_{in}$ where $C_{in}=C_{AB}+C_{HEMT}+C_{access}$. $C_{HEMT}\simeq 160$ fF is the input gate-source capacitance of the HEMT (here an ATF-34143 transistor from Agilent ).  $C_{access}\simeq 70$ fF is the access capacitance due to, at most, 1\,mm long bonding wires necessary to connect the HEMT to the sensing capacitor. Taking  $C_{in}=0.48$ pF, this gives a typical signal amplitude of $V_{s}$=0.33$\mu$V. 

Let us now discuss the timescales that are involved in the single-shot measurement of a flying qubit. 60\,ps short Lorentzian voltage pulses are routinely achieved to generate clean single charge pulses in the form of levitons, see \cite{Dubo13,Jull14,Roussely2018}. 
Using a charge propagation velocity $v=3.10^{4}$ m/s in the integer quantum Hall effect with filling factor $\nu=1$ \cite{Kuma2011}, this corresponds to a wave packet size of 1.8$\mu$m, much shorter than half of the meander length ($L/2=0.5$ mm ), a necessary condition for the charge to generate an oscillating voltage across the sensing electrodes with maximum amplitude. The oscillation frequency is expected to be $f=v/2L= 15$ MHz and the oscillations will last for the time $\tau=N 2L/v=4.5\mu$s which defines the detection time. This time also defines the maximum rate at which single-shot quantum operation must be performed.

In the following, we consider the noise performance of our single-shot detector. At 15 MHz, the ATF-34143 HEMT, biased at a moderate voltage gain has an input voltage noise of $V_{n}=$0.2 nV/Hz$^{1/2}$ at cryogenic temperature in the white noise regime.
For a detection time of 4.5$\mu$s, the r.m.s. voltage noise is $V_{Noise}=V_{n}/\sqrt\tau=$ 0.093 $\mu$V and the signal to noise ratio for single-charge detection is $S/N=3.6$. 
How good is this $S/N$ ratio in terms of readout fidelity? For a QPC set at transmission $D$=1/2 with a detection threshold for a transmitted flying qubit of $V_{s}/2$, the error probability for Gaussian noise is 
$P_{Err.}=\frac{1}{\sqrt{2\pi V_{Noise}^{2}}}\int_{V_{s}/2}^{\infty} \exp{(-U^{2}/2V_{Noise}^{2})} dU $. 
Using the complementary error function :
\begin{equation}\label{ProbaError}
    P_{Err.}=\frac{1}{2}  \operatorname{erfc}(\frac{V_{s}}{2\sqrt{2}V_{Noise}}) =3.5\% 
\end{equation} 
This gives a read-out fidelity better than 96$\%$. 

Now that we have shown the potential of our detector to perform a single-shot readout of a single-electron flying qubit, we would like to refine our understanding of the charge wavepacket propagation. 
There are two important issues. 
The first one is an accurate prediction of the charge pulse propagation velocity, the second one is about the possible spreading of the charge pulse. 
Spreading can arise from frequency dispersion and from a possible effect of the propagation discontinuity  occurring when the charge passes below the 10$\mu$m gap separating the two sensing electrodes.

The propagation of charge pulses at the periphery of a 2D electron system in the  presence of strong perpendicular magnetic field is described by the so-called edge magneto plasmons (EMP) \cite{Volk88,Alei94,John2003,Glat85,Asho92,Taly92,Zhit94,Erns97,Kamata2010}. 
Plasmons are collective excitations of electrons mediated by the Coulomb interaction. 
In two-dimension, the zero field two-dimensional plasmon  dispersion relation is $\omega_{p}(q)=\sqrt{n_{s}e^2q/2m_{eff}\epsilon\epsilon_{0}}$. In presence of perpendicular magnetic field $B$, it splits into two branches corresponding to qualitatively different modes ($n_{s}$ is the electron density, $m_{eff}$ the effective electron mass and $\epsilon$ the permittivity of the host GaAs semiconductor and $q$ the wavelength). 
One finds a bulk mode with $\omega_{p}\longrightarrow \sqrt{\omega_{p}(q)^2+\omega_{c}^2}$ where $\omega_{c}=eB/m_{eff}$ is the cyclotron frequency and a chiral edge mode with $\omega_{p}\,\longrightarrow \,\, \simeq \omega_{p}(q)^2/\omega_{c}\propto \sigma_{xy}q/\epsilon\epsilon_{0}$ where $\sigma_{xy}$ is the Hall conductance. 
The latter expression for the chiral mode dispersion relation is derived from a dimensional analysis and is only approximate. 
The exact expression contains a $ln(q)$ term due to the long-range interaction \cite{Volk88,Alei94,John2003}. 
Here we consider the strong screening regime where the distance between the sensing metallic electrodes and the 2D electron gas is $d\simeq 100$nm which is much shorter than the electron wave packet width ($\simeq 1\mu$m) and the edge channel width $a > 2\mu$m \cite{Kuma2011}. 
In this limit on has: 
\begin{equation}\label{chiraldispersion}
    \omega(q)=\sigma_{xy}q(\dfrac{1}{C}+\dfrac{1}{C_{Q}})
\end{equation}
where $C=\epsilon\epsilon_{0}a/d$ is the capacitance per unit length of the capacitor formed by the chiral channel and the metallic electrode, $C_{Q}=e^2/(hv_{D})$ is the quantum capacitance of the channel due to the Fermi statistics, $v_{D}$ the single-particle drift velocity and $\sigma_{xy}=e^2/h$ for filling factor $\nu=1$.
From Eq.\,\ref{chiraldispersion} we obtain the charge pulse propagation velocity:
\begin{equation}\label{pulsevelocity}
    v=v_{p}+v_{D}
\end{equation}
with: 
\begin{equation}\label{Coulombvelocity}
    v_{p}=\dfrac{e^{2}}{h}\dfrac{d}{\epsilon\epsilon_{0}a}
\end{equation}
the velocity of pure Coulomb origin. Taking $d=100nm$, $\epsilon=12.8$ for the GaAs host semiconductor and the edge width $a=2\mu$m measured in \cite{Kuma2011}, one finds $v_{p}=1.7$ $10^{4}$m/s. One expects the single-particle drift velocity $v_{D}=E_{conf}/B$ to be of the same order of magnitude. This is compatible with measurements in \cite{Kuma2011} where $v=3.$ $10^{4}$m/s is found  giving a single particle drift velocity $v_{D}=1.3$ $10^{4}$m/s compatible with a reasonable value of the confinement electric field $E_{conf}\simeq 50$mV/$\mu$m  in a magnetic field $B=4$ Tesla for $\nu=1$ and electron density $10^{15}/m^{2}$.

From the above discussion we see that we have a good understanding  of EMP propagation properties. The charge pulse velocity $c=3.10^{4}$m/s used for the design of the detector in Fig.2 is, therefore, a reasonable value. For a  different velocity a scaling of the geometrical design is possible. 

From the above discussion, we can also conclude that dispersion is absent, as the dispersion relation is linear with wave number $q$.
Dissipation may be another source of wavepacket dispersion but this is unlikely in the quantum Hall regime as a fine tuning of the magnetic field provides nearly zero longitudinal conductance $\sigma_{xx}$. 

Finally, we have to examine if the absence of screening in the small $10\mu$m gap between the two sensing electrodes can be a source of charge pulse dispersion. To understand this qualitatively, we will not consider a complete lack of screening in the electrode gap (which is a hard mathematical problem to solve because of long range interaction), but, we will instead consider that the distance between the chiral channel and the electrodes in the gap is large, i.e. $d(x)$ becomes spatially dependent  where $x$ is the propagation direction. Defining respectively the charge pulse density, the associated current and the local potential experienced by the propagating pulse by $\rho(x,t)$, $I(x,t)$ and $V(x,t)$, 
we have :
\begin{equation}\label{current}
    \dfrac{\partial I}{\partial x}=-\dfrac{\partial \rho}{\partial t}
\end{equation}
\begin{equation}\label{localrelation}
    \rho = C(x) V = C(x) \dfrac{I}{\sigma_{xy}}
\end{equation}
where we have used the capacitive relation, with $C(x)\propto 1/d(x)$, and use the Hall conductance relating current and voltage. Defining the, now spatially dependent, local velocity as $v(x)=\sigma_{xy}/C(x)$, we get:
\begin{equation}\label{waveequation}
    \dfrac{\partial}{\partial x}(v(x)\rho)=-\dfrac{\partial \rho}{\partial t}
\end{equation}
The solution of Eq.\ref{waveequation} is of the form:
\begin{equation}\label{solution}
    \rho (x,t) \propto \dfrac{1}{v(x)}f(t-\int_{-\infty}^{x}dx'/v(x'))
\end{equation}
with the function $f(t)$ defined so as to match the initial shape of the charge pulse entering in the first sensing electrode.

From Eq.\ref{solution}, we observe that the chirality of the propagation is preserved and no backscattering occurs. 

\section{\label{sec:level4} QND FLYING QUBIT DETECTION}

In the previous section, we have seen that the meanderline non-QND detector can detect charge pulses carrying less than a single elementary charge. 
Here we will use this property and show a direct application for QND single charge detection.
The idea is to put the meanderline detector at the output of a dc biased quantum conductor whose electronic transmission is sensitive to electronic charges. 
This could be an electronic interferometer, like a Mach-Zehnder or a Fabry-P\'erot interferometer, or a quantum point contact. 
The principle is as follows: the electron flying qubit to be QND detected passes very close to the interferometer and changes its phase during the interaction time $\tau_{int}=l_{int}/v.$ where $v$ is the velocity of the flying electron on its edge channel and $l_{int}$ is the interaction length. 
If the bias voltage $V_{dc}$ is of the order of $V_{dc}=(eV/h)\tau_{int}/\Delta T$, with $\Delta T$ the change of the  transmission during the interaction, a negative current pulse of width $\tau_{int}$ carrying a charge $Q\simeq e$ is emitted in the output channel of the interferometer and subsequently detected by the meanderline detector. 
A similar detection can be done if the detector is placed at the reflection output of the interferometer. 
In this case,  a positive pulse of the same amplitude  will be detected by the meanderline detector placed at the reflection output of the interferometer. 
Fig.\,3 shows an example where a Fabry-P\'erot interferometer formed by a quantum dot in the QHE regime is used and the meanderline detector is placed at the reflection output. A floating gate is used to mediate the electrostatic coupling between the flying qubit and the interferometer during the interaction  time $\tau_{int}$. 

With a QND detection, the electronic flying qubit remains available for further quantum manipulation, a distinct feature from the non-QND detector described in section III. The limitation is the few $\mu$s reading time which is set by the meanderline detector.

\begin{figure}
  \includegraphics[width=8.5cm,keepaspectratio=true,clip=tue]{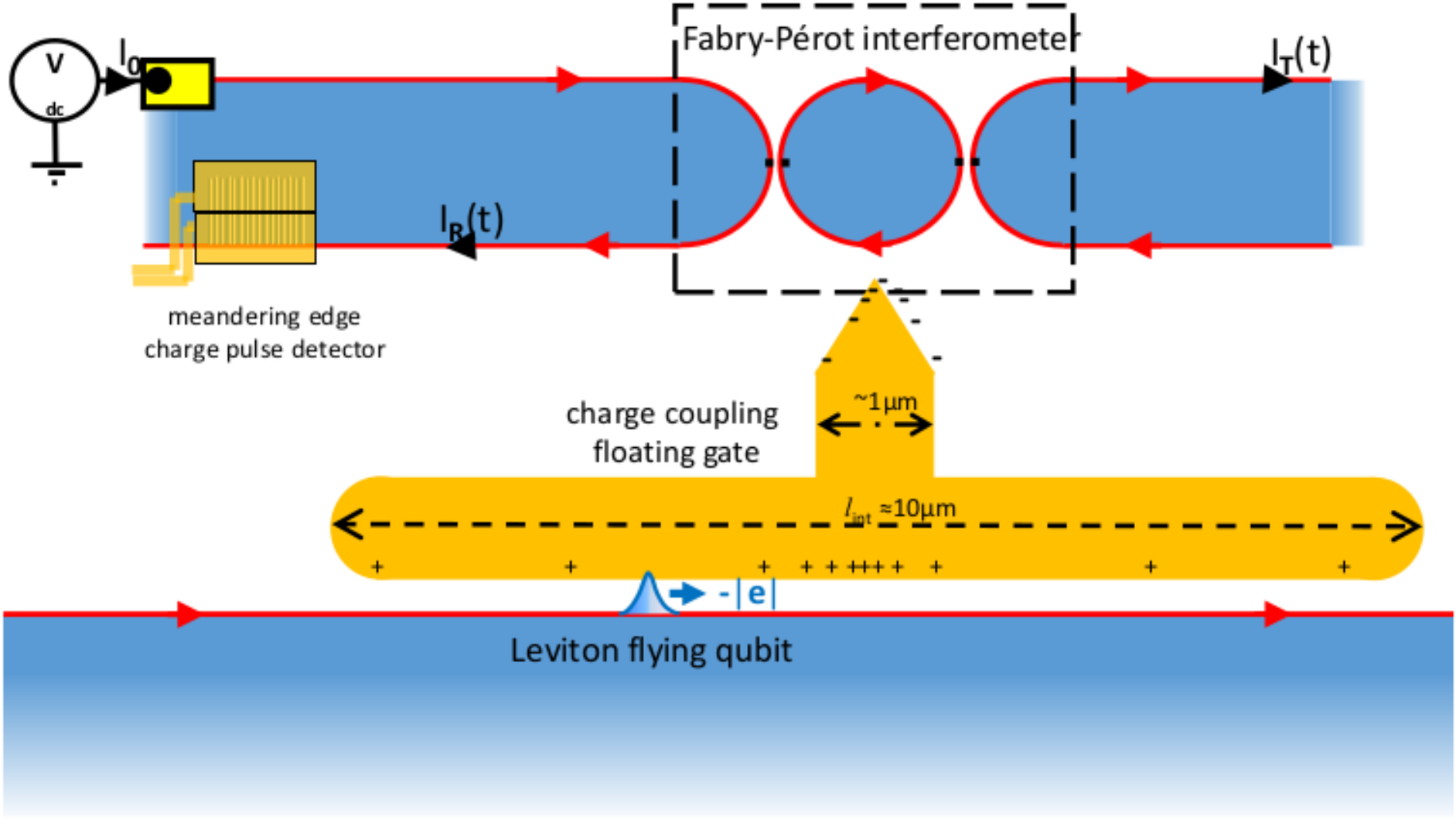}
  \caption{In-flight QND electron detection. Two separate 2 dimensional electron gases are shown in blue. The red line with arrows denotes chiral edge channels which propagate along the boundaries of the 2DEGs. The electron flying qubit to be detected, here a leviton, propagates from left to right along the edge channel of the bottom 2DEG. 
  A floating gate ensures the capacitive coupling to an electronic interferometer, here a Fabry-P\'erot made from a submicron size quantum dot connected to the left and right leads by a tunnel barrier. The gate is long enough to capture the electric field generated by the leviton for a time $\tau_{int}= l_{int}/v$ and has a sharp tip to concentrate a significant fraction of the leviton electric field on the quantum Dot. The interferometer is initially tuned to perfect transmission  and biased by the voltage $V_{dc}$. The interaction with the leviton gives rise to finite reflection during the time $\tau_{int}$ resulting in a short current pulse which is further detected by the meander edge channel charge detector. }
  \label{fig:QNDdetector}
\end{figure}

\section{\label{sec:level5} CONCLUSION}
We have proposed a single-shot readout flying qubit detector, filling an important gap in the tools required to perform \textit{Full Electron Quantum Optics}. We have shown that all parameters concerning the design of the detector are known. The principle on which the detector is based leads to a reliable estimation of its sensitivity. 
We predict an electron detection fidelity of 96\%.
Improvement is possible by changing the number N of  meander segments or their length to tune the signal frequency and the detection duration. 
By coupling such a detector to an electronic interferometer, QND detection of a flying qubit also becomes possible in a reliable way.

Realization of such detectors (QND and non-QND) will have an important impact in quantum condensed matter physics experiments dealing with electrons. 
Beyond electron-flying-qubit applications, we could think of quantum transport experiments where the current is reduced to its most elementary level: a single electron. The detector will allow to study electron full counting statistics for the first time on the particle number level, a domain where no experimental result in the full quantum regime is available, despite large theoretical work \cite{coherentstates,Hassler08,Kind03}.
Having single-shot detection will lead to an important breakthrough in the field of electronic quantum optics and beyond, in a way similar to the advent of single photon detectors in quantum optics. 
A direct application would be quantum random number generation. 
Combining a QPC with tunable transmission and single-shot detection will provide a Bernoulli factory \cite{Pate2019}. 

Most importantly, our detector is not limited to the detection of integer charge as it is not based on charge trapping which would require charge quantization.
Instead, the detector is suitable for the detection of fractional charges. 
To give an example, let us consider the fractional quantum Hall effect (FQHE) regime at filling factor $\nu=1/3$. As explained in \cite{Glat2016} and in \cite{Kapf2019} (supplementary material) elementary fractional charge pulses can be generated using levitons in the FQHE regime \cite{Rech17,Safi}. 
As $\sigma_{xy}$ is 3 times  smaller  than for the situation described above ($\nu=1$), the propagation velocity is three times smaller and consequently, the detection time is three times larger for the same geometry. 
The signal to noise ratio $V_{s}/V_{n}$ used in Eq. (1) is only reduced by a factor of  $1/\sqrt(3)$ and leads to $P_{Err.}=17\%$ for $e*=e/3$ FQHE quasiparticle detection. These numbers are encouraging to perform Hong Ou Mandel like coincidence experiment to study both,  abelian and non-abelian statistics. 

\section{Acknowledgements}
Useful discussions with the CEA Saclay Nanoelectronics team are acknowledged. We acknowledge financial support from the French National Agency (ANR) in the frame of its program FullyQuantum 16-CE30-0015 and funding from the European Union’s H2020 research and innovation programme under grant agreement No 862683 UltraFastNano.

\bibliography{References}

\end{document}